\documentclass[aps,twocolumn,showpacs,preprintnumbers,amsmath,amssymb,showpacs,showkeys]{revtex4}
\usepackage{amssymb}
\usepackage{amsmath}
\usepackage{amsfonts}
\usepackage{epsfig}
\usepackage{graphicx}
\usepackage{tabularx}
\newcommand{\seq}{\begin{subequations}}
\newcommand{\sen}{\end{subequations}}
\newcommand{\eq}{\begin{eqnarray}}
\newcommand{\en}{\end{eqnarray}}

\def\shiftdown#1{#1\llap{\lower.04ex\hbox{#1}}}

\newcommand{\bfq}{{\bf q}_{\perp}}

\newcommand{\bfk}{{\bf k}_{\perp}}

\begin{document}

\title{Nucleon GPDs in a light-front quark model \\
derived from soft-wall AdS/QCD}

\author{Alfredo Vega$^1$,
        Ivan Schmidt$^2$,
        Thomas Gutsche$^3$,
        Valery E. Lyubovitskij$^3$\footnote{On leave of absence 
from Department of Physics, Tomsk State University, 634050 Tomsk, Russia}
\vspace*{1.2\baselineskip}\\
}

\affiliation{
$^1$ Departamento de F\'isica y Astronom\'ia,
     Universidad de Valpara\'iso,\\
     Avenida Gran Breta\~na 1111, Valpara\'iso, Chile
\vspace*{1.0\baselineskip} \\
$^2$ Departamento de F\'\i sica y Centro Cient\'i fico Tecnol\'ogico 
de Valpara\'\i so (CCTVal), Universidad T\'ecnica Federico Santa Mar\'\i a, 
Casilla 110-V, Valpara\'\i so, Chile
\vspace*{1.2\baselineskip} \\
$^3$ Institut f\"ur Theoretische Physik, Universit\"at T\"ubingen, \\
Kepler Center for Astro and Particle Physics,
\\ Auf der Morgenstelle 14, D-72076 T\"ubingen, Germany
\vspace*{1.2\baselineskip}\\
}

\date{\today}

\begin{abstract}

We study the helicity-independent generalized parton distributions 
of nucleons in the zero skewness case, based on a particular light-front 
quark model derived in a soft-wall AdS/QCD approach.

\end{abstract}



\maketitle

\section{Introduction}

Generalized parton distributions (GPDs) are important objects 
containing essential information about the hadronic 
structure~\cite{Mueller:1998fv}-\cite{Diehl:2013xca}.
Unfortunately, given the non-perturbative nature of these functions, 
it is not possible to calculate them directly from 
Quantum Chromodynamics (QCD), and this situation has motivated the 
development of other ways to access the GPDs:  
namely, extraction from the experimental measurements of hard processes,
direct calculation using lattice QCD, and different phenomenological models. 
The last procedure is based on parametrizations of the quark wave function 
or directly the GPDs, using constraints imposed by sum rules, which relate 
the parton distribution functions to nucleon electromagnetic 
form factors~\cite{Ji:1996ek,Radyushkin:1997ki}, or including a precise 
$x$ behavior to improve the calculations of some hadron properties starting 
from QPDs. Some examples of this procedure can be found e.g
in~\cite{Burkardt:2002hr}-\cite{Diehl:2013xca}.

Within the phenomenological models used recently some  are based on the 
gauge/gravity duality, and are in general called holographic AdS/QCD models. 
They suppose the existence of a gravity theory dual to QCD, and are divided 
into two classes, the top-down approach (where we start from a string theory 
leading to a low energy gauge theory with some QCD properties) and the 
bottom-up models (phenomenological approach where the geometry of 
an AdS space and bulk fields  are specified in order to incorporate some 
basic properties of QCD). In turn, these last ones are divided into 
hard wall~\cite{Polchinski:2001tt,deTeramond:2005su} and soft wall 
models~\cite{Karch:2006pv,Brodsky:2007hb,Forkel:2007cm,%
Branz:2010ub,Gutsche:2011vb}, depending on the way conformal invariance 
is broken on the AdS side.

The bottom-up soft wall models have proven to be quite useful because of 
their simplicity and variety of successful applications. For example, 
they have been used for deep inelastic scattering~\cite{Polchinski:2002jw,%
BallonBayona:2007qr,Pire:2008zf,Braga:2011wa}, hadronic mass 
spectra~\cite{deTeramond:2005su,Forkel:2007cm,Gutsche:2011vb,%
Vega:2008af,Vega:2008te}, hadronic couplings and models with chiral 
symmetry breaking~\cite{DaRold:2005zs,Erlich:2005qh,Vega:2010ne,%
Vega:2011tg,Gutsche:2012ez}, quark potentials~\cite{Andreev:2006ct,%
Jugeau:2008ds}, calculations of hadronic form 
factors~\cite{Brodsky:2007hb,HFF1,HFF2,Gutsche:2012bp},  generalized parton 
distributions (GPD)~\cite{Vega:2010ns,Vega:2012iz,Nishio:2011xa}, 
light front wave functions (LFWF)~\cite{Brodsky:2006uqa,%
Brodsky:2007hb,Vega:2009zb,Branz:2010ub}, etc.

In Refs.~\cite{Vega:2010ns,Gutsche:2012bp} we applied the AdS/QCD 
correspondence to the nucleon GPDs and electromagnetic form factors, 
using both soft-wall and hard-wall holographical models. In particular, 
the soft-wall version qualitatively reproduced the small and large $x$ 
and $Q^2$ behavior of the GPDs. In the case of related quantities --- 
nucleon form factors --- the following results were achieved: 
i) we analytically reproduced their scaling at large values of Euclidean 
momentum squared $Q^2$, consistent with quark counting 
rules~\cite{Brodsky:1973kr}; ii) we systematically included the 
contributions of higher Fock states; iii) a reasonable description 
of hadronic factors and their ratios at both small and large~$Q^2$ 
was obtained. In Ref.~\cite{LFWF_paper} we proposed a light-front 
quark model based on phenomenological LFWF for the nucleon, which 
is motivated in a soft-wall AdS/QCD approach. We showed that this 
model produces the parton distributions and hadronic form factors 
consistent with quark counting rules and the Drell-Yan-West (DYW) 
duality~\cite{Drell:1969km}. We showed that the inclusion of the 
effects of the longitudinal wave function of nucleons is sufficient 
to get consistency with model-independent scaling laws. The main 
objective of this paper is to give an analysis of how to improve 
the description of quark distributions in the nucleon by including 
the effect of the longitudinal wave function of nucleons.

The paper is structured as follows. In Sec.~II we discuss our approach. 
In Sec.III we present the numerical analysis of the quark distribution 
functions, magnetic densities and GPDs. Finally in Sec.~IV we present 
our conclusions.

\section{Framework}

We start by writing down the relations  ~\cite{Radyushkin:1998rt} 
between the nucleon Dirac $F_1^N$ and Pauli $F_2^N$ form factors, 
the form factors of valence quarks in nucleons ($F_1^q$ and $F_2^q$, 
$q=u,d$), and the valence quark GPDs ($\mathcal{H}^{q}$ and 
$\mathcal{E}^{q}$):

\eq\label{FF}
F_i^{p(n)}(Q^2) &=& \frac{2}{3} F_i^{u(d)}(Q^2)
                 -  \frac{1}{3} F_i^{d(u)}(Q^2)
\en
and
\eq
F_1^q(Q^2) &=& \int_{0}^{1} dx \, \mathcal{H}^{q}(x,Q^2)  \,,\\
F_2^q(Q^2) &=& \int_{0}^{1} dx \, \mathcal{E}^{q}(x,Q^2)  \,.
\en

At $Q^2=0$ the GPDs $\mathcal{H}^{q}$ and $\mathcal{E}^{q}$ reduce to 
the valence $q_v(x)$ and magnetic $\mathcal{E}^{q}(x)$ quark densities

\eq\label{CondicionDensidadQuark}
\mathcal{H}^{q}(x,0)=q_{v}(x)\,,
\quad
\mathcal{E}^{q}(x,0)=\mathcal{E}^{q}(x) \,,
\en
which are normalized to the number of valence $u$ and $d$ quarks in the 
proton (in the case of $q_v$ distributions) and to the anomalous magnetic 
moment of the quark (in the case of $\mathcal{E}^{q}$ distributions), 
respectively:

\eq\label{normalization}
\int\limits_0^1 dx \, u_v(x) = 2\,, \
\int\limits_0^1 dx \, d_v(x) = 1\,, \
\kappa_q = \int\limits_0^{1} dx \, \mathcal{E}^q(x)\,.
\en

The constants $\kappa_q$ are related to the anomalous magnetic moments 
of nucleons $k_N = F_{2}^{N}(0)$~\cite{Lu:2007}:

\eq\label{CondicionKappa}
\kappa_{u} &=& 2 \kappa_{p} + \kappa_{n} = 1.673  \,,
\nonumber\\
\kappa_{d} &=& \kappa_{p} + 2 \kappa_{n} = -2.033 \,.
\en

In Ref.~\cite{LFWF_paper} we proposed a light-front quark model for 
nucleons which is motivated by the soft-wall AdS/QCD approach developed 
in Refs.~\cite{Vega:2010ns,Gutsche:2012bp} and is consistent with quark 
counting rules and DYW duality~\cite{Drell:1969km}. In particular, 
we showed that the quark GPDs in the nucleon are given in the 
form~\cite{LFWF_paper}:

\eq
{\cal H}^{q}(x,Q^2) &=& q_v(x) \, f(x,Q^2) \,,
\label{HqAdS}\\
{\cal E}^{q}(x,Q^2) &=& {\cal E}^q(x) \, f(x,Q^2) \,,
\label{EqAdS}
\en
where the quark densities at large $x \to 1$ behave in accordance 
with the scaling laws

\eq
q_v(x) \sim (1-x)^3\,, \quad {\cal E}^q(x) \sim (1-x)^5 \,.
\en

The function $f(x,Q^2)$ is universal for both types of quark densities 
and as in~\cite{Burkardt:2002hr}-\cite{Selyugin:2009ic} it contains 
both the $Q^2$ and $x$ dependencies.

In agreement with Refs.~\cite{Burkardt:2002hr}-\cite{Selyugin:2009ic}, 
here we do not factorize the $x$ and $Q^2$ dependencies in $f(x,Q^2)$.  
We can fulfill all constraints imposed by the small and large $x$-behavior 
and to implement the DYW duality by taking~\cite{LFWF_paper}:

\eq
f(x,Q^2) = \exp\left[- \frac{Q^2}{4\kappa^2} \, \log(1/x) (1-x)\right] \,,
\en
where $\kappa$ is the scale parameter.

As  in Ref.~\cite{Guidal:2004nd}, our function $f(x,Q^2)$ has a clear 
interpretation in the framework of the light-front formalism~\cite{LFQCD}, 
since it corresponds to the modified Regge ansatz proposed in 
Ref.~\cite{Burkardt:2002hr}. The Dirac and Pauli quark form factors 
are obtained in terms of LFWFs  by

\eq\label{FF_LFQCD}
F_1^q(Q^2) &=&
\int\limits_0^1 dx
\int\frac{d^2\bfk}{16\pi^3} \,
\biggl[
\psi_{+q}^{+\, \ast}(x,\bfk')\psi_{+q}^+(x,\bfk)
\nonumber\\
&+&\psi_{-q}^{+\, \ast}(x,\bfk')\psi_{-q}^+(x,\bfk)
\biggr] \,, \nonumber\\
& &\\
F_2^q(Q^2) &=& - \frac{2M_N}{q^1-iq^2}
\int\limits_0^1 dx
\int\frac{d^2\bfk}{16\pi^3} \nonumber\\
&\times&
\biggl[
\psi_{+q}^{+\, \ast}(x,\bfk')\psi_{+q}^-(x,\bfk)
\nonumber\\
&+&\psi_{-q}^{+\, \ast}(x,\bfk')\psi_{-q}^-(x,\bfk)
\biggr] \,. \nonumber
\en
where $\bfk' = \bfk  + (1-x) \bfq$.   In these equations $M_N$ is the 
nucleon mass, $\psi_{\lambda_q q}^{\lambda_N}(x,\bfk)$ are the LFWFs 
with specific helicities of the nucleon $\lambda_N  = \pm$ and struck 
quark $\lambda_q = \pm $, where plus and minus correspond to $+\frac{1}{2}$ 
and $-\frac{1}{2}$, respectively. Working in the frame $q=(0,0,\bfq)$ 
we have that $-q^2 = Q^2 = \bfq^2$.

The LFWFs $\psi_{\lambda_q q}^{\lambda_N}(x,\bfk)$ are defined 
as~\cite{LFWF_paper}

\eq
\psi_{+q}^+(x,\bfk) &=& \frac{m_{1q} + xM_N}{x} \, \varphi_q(x,\bfk)
\,, \nonumber\\
\psi_{-q}^+(x,\bfk) &=& -\frac{k^1 + ik^2}{x}
\, (1-x) \, \mu_q \, \varphi_q(x,\bfk)
\,, \nonumber\\
&&\\
\psi_{+q}^-(x,\bfk) &=& \frac{k^1 - ik^2}{x}
\, (1-x) \, \mu_q \, \varphi_q(x,\bfk)
\,, \nonumber\\
\psi_{-q}^-(x,\bfk) &=& \frac{m_{1q} + xM_N}{x} \, \varphi_q(x,\bfk)
\,, \nonumber
\en
where $m_{1q}$ is the mass of the struck quark. The wave function 
$\varphi_q(x,\bfk)$ can be taken from recent AdS / QCD work,  and is 
given by the product of transverse and longitudinal wave 
functions~\cite{Gutsche:2012ez,LFWF_paper}

\eq\label{LFWF_fin}
\varphi_q(x,\bfk) &=& N_q \, \sqrt{\log(1/x)} \, x^{\beta_{1q}} \,
(1-x)^{\beta_{2q} + 1}
\nonumber\\[1mm]
&\times& \exp\biggl[- \frac{{\cal M}^2}{2\kappa^2} \, x \, \log(1/x) \biggr]
\,.
\en

The invariant mass is

\eq
{\cal M}^2 = {\cal M}^2_0 + \frac{\bfk^2}{x (1-x)}
 = \frac{\bfk^2 + m_1^2}{x} \, + \, \frac{\bfk^2 + m_2^2}{1-x}
\en
and
\eq
\hspace*{-.6cm} 
N_q &=& \frac{4 \pi\sqrt{n_q}}{\kappa M_N}
\biggl[ \int\limits_0^1 dx \, x^{2\beta_{1q}}
(1-x)^{3+2\beta_{2q}} \nonumber\\
&\times&
R_q(x) \, e^{-\frac{{\cal M}_0^2}{\kappa^2} \, x \, \log(1/x)} 
\biggr]^{-1/2}   \,, \nonumber\\
\hspace*{-.6cm}
R_q(x) &=& \Big(1 + \frac{m_{1q}}{x M_N}\Big)^2
+  \frac{\kappa^2 \mu_q^2}{M_N^2} \frac{(1-x)^3}{x^2 \, \log(1/x)}
\en
is the normalization constant and $\beta_{1q}$ and $\beta_{2q}$ are 
the flavor dependent parameters  (quark masses). The parameters 
$\mu_q$ $(q=u,d)$ are fixed through the nucleon magnetic moments.

In the LFWF $\psi_{\mp q}^{\pm}(x,\bfk)$ we included the extra factor 
$(1-x)$ in order to generate an extra power $(1-x)^2$ in the helicity-flip 
parton density ${\cal E}^q(x) \sim (1-x)^5$ in comparison to the helicity 
non-flip density $q_v(x) \sim (1-x)^3$. 

\section{Parton distributions in nucleons}

\begin{figure*}[t]
  \begin{tabular}{c}
    \includegraphics[width=3.4 in]{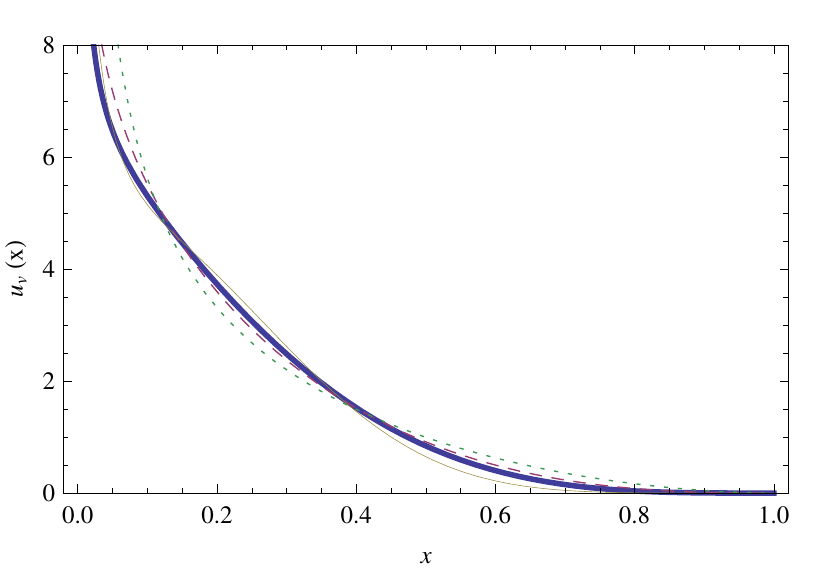}
    \includegraphics[width=3.4 in]{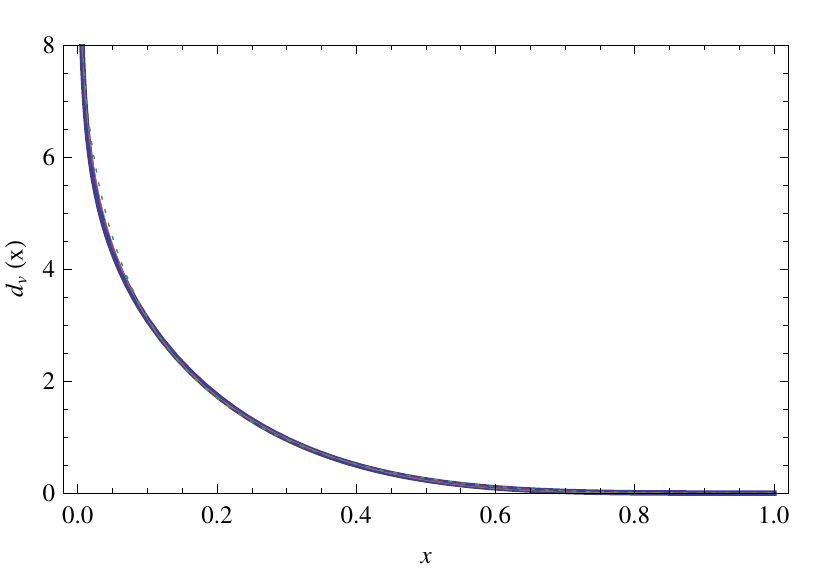}
  \end{tabular}
\caption{Comparison between parton distributions. The thick continuous 
line corresponds to the MRST global NNLO fit at the scale 
$\mu^{2} = 1~$GeV$^{2}$~\cite{Martin:2002dr}, 
while the dashed line corresponds to the 
parton distributions given using Fit I, the thin continuous line 
correspond to Fit II and the dotted line is for Fit III. Parameters 
used are summarized in Tables I and II.}

  \begin{tabular}{c}
    \includegraphics[width=3.4 in]{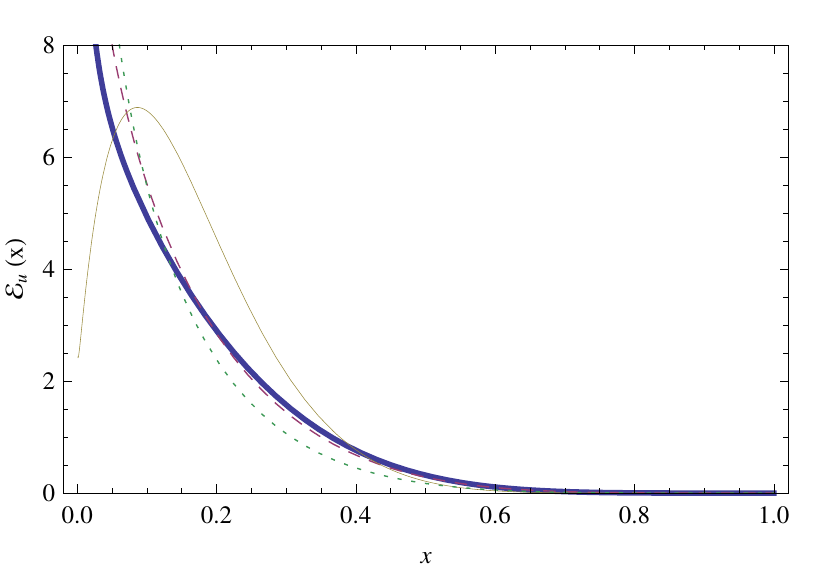}
    \includegraphics[width=3.4 in]{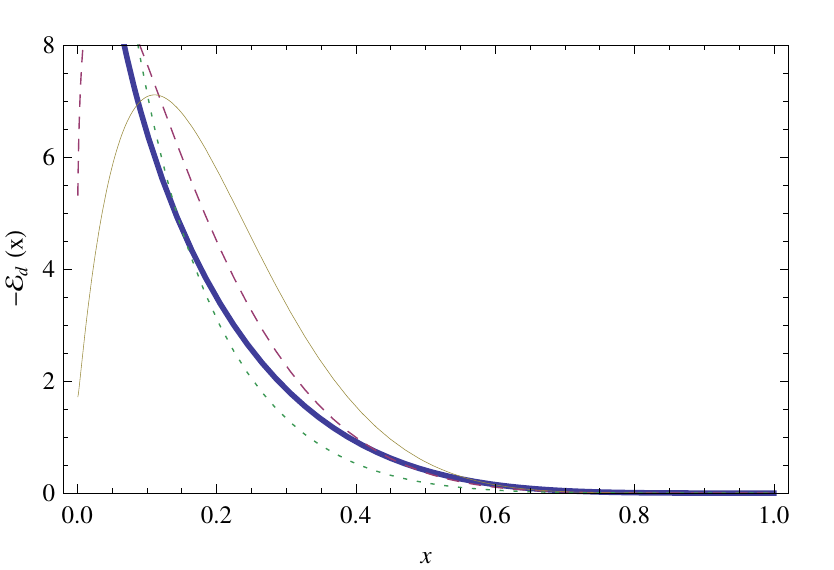}
  \end{tabular}
\caption{Comparison between parton distributions. The thick continuous 
line corresponds to the form suggested in \cite{{Diehl:2005wq}} with 
the MRST global NNLO fit at the scale 
$\mu^{2} = 1~$GeV$^{2}$~\cite{Martin:2002dr}, while the 
dashed line corresponds to the parton distributions given by Fit I, the 
thin continuous line correspond to Fit II and the dotted line is for 
Fit III. The parameters used here are indicated in Tables I and II.}
\end{figure*}

\begin{table}[b]
\begin{center}
\caption{Parameters used for three different fits in the parton
distributions $q_{u}(x)$. In all cases we consider $\kappa = 350$~MeV.}
\begin{tabular}{ c c c | c c c | c c c | c c c | c c c | c c c }
  \hline
  \hline
  & Model & & & $m_{q}$ [MeV] & & & $m_{D}$ [MeV] & & & A & & &
$\rho_{1}$ & & & $\rho_{2}$ & \\
  \hline
  & I & & & 0 & & & 0 & & & 4.76 & & & -0.18 & & & 2.56 & \\
  & II & & & 7 & & & 100 & & & 105.14 & & & 1.41 & & & 5.93 & \\
  & III & & & 300 & & & 600 & & & 63.67 & & & 1.15 & & & 2.74 & \\
  \hline
  \hline
\end{tabular}
\end{center}
\begin{center}
\caption{Parameters used for three different fits in the parton distributions 
$q_{d}(x)$. In all cases we consider $\kappa = 350$~MeV.}
\begin{tabular}{ c c c | c c c | c c c | c c c | c c c | c c c }
  \hline
  \hline
  & Model & & & $m_{q}$ [MeV] & & & $m_{D}$ [MeV] & & & A & & & $\rho_{1}$ 
  & & & $\rho_{2}$ & \\
  \hline
  & I & & & 0 & & & 0 & & & 2.87 & & & -0.21 & & & 3.76 & \\
  & II & & & 7 & & & 100 & & & 33.06 & & & 1.63 & & & 5.71 & \\
  & III & & & 300 & & & 600 & & & 27.30 & & & 1.10 & & & 3.42 & \\
  \hline
  \hline
\end{tabular}
\end{center}
\end{table}

Our results for the quark Dirac and Pauli form factors are~\cite{LFWF_paper}
\eq
F_1^q(Q^2) &=& C_q \, \int\limits_0^1 \!dx \,
x^{2\beta_{1q}} \, (1 - x)^{3+2\beta_{2q}} \, R_q(x,Q^2)
\nonumber\\
&\times&
\exp\biggl[-\frac{Q^2}{4\kappa^2} \, \log(1/x) \, (1-x)\biggr]
\nonumber\\
&\times&
\exp\biggl[ - \frac{{\cal M}_0^2}{\kappa^2} \, x \, \log(1/x) \biggr]  \,,
\nonumber\\
& &\\
F_2^q(Q^2) &=& C_q \int\limits_0^1 \!\frac{2dx}{x} \, \mu_q
\biggl(1 + \frac{m_{1q}}{x M_N}\biggr) \,
x^{2\beta_{1q}} \, (1 - x)^{5+2\beta_{2q}} \nonumber\\
&\times&
\exp\biggl[-\frac{Q^2}{4\kappa^2} \, \log(1/x) \, (1-x)\biggr]
\nonumber\\
&\times&
\exp\biggl[ - \frac{{\cal M}_0^2}{\kappa^2} \, x \, \log(1/x) \biggr]  \,,
\nonumber
\en
where
\eq
\hspace*{-.6cm}
C_q &=& N_q^2  \, \Big(\frac{\kappa M_N}{4 \pi}\Big)^2  \nonumber\\
\hspace*{-.6cm}
    &=& n_q \biggl[ \int\limits_0^1 \!dx  x^{2\beta_{1q}}
(1-x)^{3+2\beta_{2q}} \, R_q(x) \,
e^{ - \frac{{\cal M}_0^2}{\kappa^2}x \log(1/x)} \biggr]^{-1}   \!\,, \nonumber\\
& &R_q(x,Q^2) = R_q(x)
           - \frac{Q^2}{4M_N^2} \frac{\mu_q^2 \, (1-x)^4}{x^2} \,.
\en

It means that the quark densities are given by
\eq
\hspace*{-.8cm}
q_v(x) &=&  C_q \,
x^{2\beta_{1q}} \, (1 - x)^{3+2\beta_{2q}} \, R_q(x) \,
e^{ - \frac{{\cal M}_0^2}{\kappa^2} x \log(1/x)}, \\
\hspace*{-.8cm}
{\cal E}^q(x) &=& C_q \,
x^{2\beta_{1q}} \, (1 - x)^{5+2\beta_{2q}} \, P_q(x) \,
e^{ - \frac{{\cal M}_0^2}{\kappa^2}x \log(1/x)} \,,
\en
where
\eq
P_q(x) = \frac{2 \mu_q}{x} \,
\biggl(1 + \frac{m_{1q}}{x M_N} \biggr) \,.
\en

We remind again that the parameters $\beta_{1q}$ and $\beta_{2q}$ define 
the flavor structure of the longitudinal part of the hadronic LFWF. 
In Ref.~\cite{Gutsche:2012ez} we showed that the longitudinal part of 
the hadronic LFWF is an essential ingredient  for generating the mechanism 
of explicit breaking of chiral symmetry in the sector of light quarks and 
imposing the constraints required by the heavy quark symmetry. See also 
other recent papers~\cite{Forshaw:2012im,Chabysheva:2012fe} discussing 
the role of the longitudinal LFWF. Therefore, it is interesting to study 
the sensitivity of the quark densities to the choice of the parameters 
$\beta_{iq}$ and compare them with global fits.

We consider three models with usually assigned values for the quark 
$(m_q)$ and diquark $(m_D)$ masses. In model I we consider the case 
without massive quarks, in model II we consider a current quark and 
in model III we use a constituent quark mass.

We start with the analysis of the fit I
\eq\label{DensidadPartonesHolografica}
q_{v}(x) &=& A_q \, x^{\rho_{1q}} (1-x)^{\rho_{2q}} \,, \\
{\cal E}_q(x) &=&  D_q \, x^{\rho_{1q}} (1-x)^{2 + \rho_{2q}} \,,
\en
where $A_q$ and $D_q$ are the normalization constants fixed from the  
conditions~(\ref{normalization}).

The sets of free parameters $\rho_{iq}$ are fixed by comparison with the 
parton densities that consider the global fit of 
MRST2002~\cite{Martin:2002dr}. Notice that in the literature there exist 
several alternative parametrizations for the quark distribution functions, 
see e.g. Refs.~\cite{Guidal:2004nd,Selyugin:2009ic}.

For models II and III we use
\eq
\label{DensidadPartonesHolograficaMasa1}
\hspace*{-.6cm}
q_{v}(x) &=& A_q \, x^{\rho_{1q}} (1-x)^{\rho_{2q}} R_q(x) \,
e^{ - \frac{{\cal M}_0^2}{\kappa^2} x \log(1/x)}, \\
\label{DensidadPartonesHolograficaMasa2}
\hspace*{-.6cm}
{\cal E}_q(x) &=&  D_q \, x^{\rho_{1q}} (1-x)^{2 + \rho_{2q}} P_q(x) \,
e^{ - \frac{{\cal M}_0^2}{\kappa^2} x \log(1/x)}\,.
\en
The parameters for each fit are summarized in Tables I and II. 
In Figs. 1 and 2 we show the parton distributions calculated with these 
values. For all cases we use $\kappa = 350$ MeV as in \cite{Vega:2010ns}.

As for example in \cite{Guidal:2004nd,Selyugin:2009ic} a standard 
representation of ${\cal E}_q(x)$ is
\eq
\label{ComparacionE1}
{\cal E}_u(x) &=& \frac{k_{u}}{N_{u}} (1 - x)^{\kappa_{1}} u(x)\, \\
\label{ComparacionE2}
{\cal E}_d(x) &=& \frac{k_{d}}{N_{d}} (1 - x)^{\kappa_{2}} d(x)\,,
\en
where $\kappa_{1}=1.53$ and $\kappa_{2}=0.31$, and according to the 
normalization
\eq
\hspace*{-.3cm}
k_{u}=1.673,~~k_{d}=-2.033,~~N_{u}=1.53
,~~N_{d}=0.946.
\en

We compare our results to these expressions for ${\cal E}_q(x)$. 
Fig. 2 summarizes our results and the comparison for the magnetic densities.

\section{Conclusions}

Starting from a light-front quark model we derived nucleon PDFs and GPDs 
consistent with scaling rules. Then we gave a numerical analysis of quark 
PDFs for three parameter sets. In version I we considered the holographical 
model without including massive quarks. In models II and III quark masses, 
current or constituent ones, are included.

In our expressions for $q(x)$ we obtain a good representation in each 
parameter version, as evident from Fig. 1.  For the ${\cal E}_q(x)$ of 
Fig. 2 the situation is different, we get agreement with the standard 
representations of Eqs. (\ref{ComparacionE1}) and (\ref{ComparacionE2}) 
just in model III.

If we consider in the expression of (\ref{DensidadPartonesHolograficaMasa2}) 
an arbitrary index $\sigma_{2}$ (instead of $2+\rho_{2}$) the agreement 
with ${\cal E}_q(x)$ can be improved but at the cost of a new parameter.

\begin{acknowledgments}

This work was supported by the DFG under Contract No. LY 114/2-1, 
by FONDECYT (Chile) under Grant No. 1100287 and by CONICYT (Chile) 
under Grant No. 7912010025. The work is done partially under the 
project 2.3684.2011 of Tomsk State University. V. E. L. would like 
to thank Departamento de F\'\i sica y Centro Cient\'\i fico 
Tecnol\'ogico de Valpara\'\i so (CCTVal), Universidad T\'ecnica 
Federico Santa Mar\'\i a, Valpara\'\i so, Chile for warm hospitality.

\end{acknowledgments}

\end{document}